\documentclass[12pt]{iopart}

\usepackage{graphicx}
\usepackage{hyperref}
\usepackage{amssymb}
\usepackage{xcolor}
\begin{document}

\title[Search for echoes on the edge of quantum black holes]{Search for echoes on the edge of quantum black holes}

\author{Jahed Abedi}

\address{Beijing Institute of Mathematical Sciences and Applications (BIMSA), Huairou District, Beijing 101408, P. R. China}
\address{Department of Mathematics and Physics, University of Stavanger, NO-4036 Stavanger, Norway}
\address{\small{Albert-Einstein-Institut, Max-Planck-Institut f{\"u}r Gravitationsphysik, Callinstra{\ss}e 38, 30167 Hannover, Germany}}
\address{\small{Leibniz Universit{\"a}t Hannover, 30167 Hannover, Germany}}
\ead{jahed.abedi@bimsa.cn}

\begin{abstract}
I perform a template-based search for stimulated emission of Hawking radiation (or Boltzmann echoes) by combining the gravitational wave data from 47 binary black hole merger events observed by the LIGO-Virgo-KAGRA collaboration. With a Bayesian inference approach, I found no statistically significant evidence for this signal in either of the 3 Gravitational Wave Transient Catalogs GWTC-1, GWTC-2 and GWTC-3. While the data does not provide definitive evidence against the presence of Boltzmann echoes, the Bayesian evidence for most events falls within the range of 0.3–1.6, with the hypothesis of a common (non-vanishing) echo amplitude for all mergers being weakly disfavoured at 2:5 odds.
The only exception is GW190521, the most massive and confidently detected event ever observed, which shows a positive evidence of 9.2 for stimulated Hawking radiation. The ``look-elsewhere'' effect for this outlier event is assessed by applying two distinct methods to add simulated signals in real data, before and after the event, giving false (true) positive detection probabilities for higher Bayes factors of $1.5^{+1.2}_{-0.9}\%$, $4.4^{+2.0}_{-2.0} \%$  ($35 \pm 7 \%$, $35 \pm 15 \%$). An optimal combination of posteriors yields an upper limit of  $A < 0.4$ (at 90\% confidence level) for a universal echo amplitude, whereas $A \sim 1$ was predicted in the canonical model. To ensure the robustness of the results, I have employed an additional method to combine the events hierarchically. This approach involves using a target gaussian distribution and extracting the parameters from multiple uncertain observations, which may be affected by selection biases.
\end{abstract}

\section{Introduction}
Post-merger gravitational wave (GW) echoes are our most direct observational windows into the quantum structure of black hole (BH) event horizons \cite{Cardoso:2016rao,Cardoso:2016oxy,Abedi:2020ujo}, while their non-existence would rule out different hypotheses about the nature of these enigmatic objects. The best view of these horizons can be achieved by combining a large number of binary black hole (BBH) merger events.  As such, the GW data release for BBH mergers during the first, second and third observing run of LIGO/Virgo observatories \cite{LIGOScientific:2018mvr,LIGOScientific:2020ibl,LIGOScientific:2021djp,Abbott:2016blz,TheLIGOScientific:2016src,Abbott:2017vtc,Abbott:2018lct} provides an unprecedented opportunity to test classical general relativity (GR), as well as its alternatives, in the strong gravity regime. One can assume GR as the base model and contrast it to GR+phenomenological echo waveforms to see which one fits the data better. Nonetheless, despite many attempts in searching for echoes \cite{Abedi:2016hgu,Conklin:2017lwb, Abedi:2018npz,Uchikata:2019frs, Holdom:2019bdv,Westerweck:2017hus,Nielsen:2018lkf,Salemi:2019uea,Lo:2018sep,Tsang:2019zra,Abbott:2020jks,Wang:2020ayy,Westerweck:2021nue,Ren:2021xbe,LIGOScientific:2021sio,Abedi:2021tti,Conklin:2021cbc}  using different models, we still lack a waveform as physical/accurate as waveforms in GR. Additionally, there is no consensus on the optimal procedure to combine the events, for best sensitivity to fundamental physics. Although, the reported GW detections have so far been consistent with predictions of GR \cite{TheLIGOScientific:2016src,Abbott:2017vtc,Abbott:2018lct}, the first search for echoes \cite{Abedi:2016hgu}, motivated by a resolution to the BH quantum information paradox, was conducted for the first observing run of the Advanced LIGO detectors (O1), which then motivated further searches within different GW data analysis frameworks, and using more physical echo waveforms. Accordingly, several attempts with replication and extention were made with positive \cite{Abedi:2016hgu, Conklin:2017lwb, Abedi:2018npz,Uchikata:2019frs,Holdom:2019bdv,Conklin:2021cbc}, mixed \cite{Westerweck:2017hus,Nielsen:2018lkf,Salemi:2019uea}, and negative \cite{Lo:2018sep,Uchikata:2019frs,Tsang:2019zra,Abbott:2020jks,Wang:2020ayy,Miani:2023mgl,Uchikata:2023zcu} results. These searches lead to tentative evidence and detection found with different groups \cite{Abedi:2016hgu,Conklin:2017lwb,Westerweck:2017hus,Abedi:2018npz,Salemi:2019uea,Uchikata:2019frs,Holdom:2019bdv,Abedi:2021tti,Conklin:2021cbc} at false alarm rates of $0.002\%-5\%$ (but see \cite{Westerweck:2017hus,Ashton:2016xff,Abedi:2017isz,Abedi:2018pst,Salemi:2019uea,Abedi:2020sgg,Abedi:2020ujo} for the ongoing discussion, comments, and rebuttals on statistical significance of these findings that motivate further investigations). So far, the searches for echoes have employed three strategies that can be classified into:
\begin{enumerate}
\item Waveform dependent \cite{Abedi:2016hgu,Westerweck:2017hus,Nielsen:2018lkf,Lo:2018sep,Uchikata:2019frs,Abbott:2020jks,Wang:2020ayy,Westerweck:2021nue,LIGOScientific:2021sio,Abedi:2021tti,Uchikata:2023zcu}.
\item Model-agnostic or coherent \cite{Abedi:2018npz,Conklin:2017lwb,Salemi:2019uea,Holdom:2019bdv,Ren:2021xbe,Abedi:2021tti,Conklin:2021cbc,Miani:2023mgl}.
\item Electromagnetic confirmation by Gill et al. \cite{Gill:2019bvq}.
\end{enumerate}
For more details, discussions and review please see \cite{Abedi:2020ujo}.
\begin{figure*}[ht]
\centering
\includegraphics[width=1\textwidth]{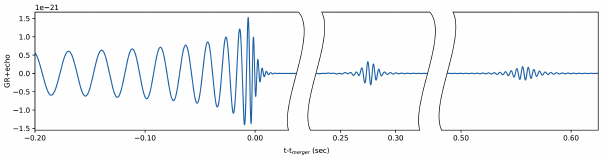}
\caption{Boltzmann GW echoes template for GW150914 like signal with amplitude $A=1$. }
\label{echo_pic_3}
\end{figure*}

A confirmed detection of echoes would  imply that the BH horizon is not totally absorbing. This would lead to post-merger repeating signals which are produced in the cavity that traps GWs between the classical angular momentum barrier and the near-horizon membrane/firewall \cite{Abedi:2016hgu,Cardoso:2016rao,Cardoso:2016oxy,Wang:2018gin,Abedi:2021tti}. However, firewalls are not a necessary condition to have observable echoes \cite{Abedi:2022omf}. Stimulated emission of Hawking radiation, caused by the GWs that excite the quantum BH microstructure has a similar effect  \cite{Oshita:2019sat,Wang:2019rcf,Xin:2021zir,Srivastava:2021uku,Abedi:2021tti}. As shown in Fig. \ref{echo_pic_3}, the trapped GWs slowly leak out, leading to repeating echoes within time intervals of:
\begin{eqnarray}
\Delta t_{\rm echo} \simeq \frac{4 G M_{\rm BH}}{c^3}\left(1+\frac{1}{\sqrt{1-a^2}}\right) \times\ln\left(\frac{G M_{\rm BH}}{ c^2 \ell_{\rm QG}}\right),\label{delay}
\end{eqnarray}
where $M_{\rm BH}$ and $a$ are the mass and the dimensionless spin of the final BH remnant. Here, $\ell_{\rm QG}$ is the characteristic  physical length scale for quantum gravity effects where GWs are reflected near the (would-be) horizon. For $\ell_{\rm QG}=\ell_{\rm Planck}$, the reflection happens at a Planck distance from the horizon. More generally, for $\ell_{\rm QG}= \ell_{\rm Planck}/\Lambda$, deviations from GR happen sub-Planckian $\Lambda>1$ or super-Planckian $\Lambda<1$ scales. 

Here, in comparison to former attempts, I used a more physical waveform, based on stimulated Hawking radiation\cite{Oshita:2019sat,Wang:2019rcf} to test for the existence of echoes. Furthermore, I adopt the Bayesian methodology and p and model, as in Abedi et al. \cite{Abedi:2021tti}.

In this approach of combining events I assume echo model is the same for all the events. In particular, I assume all the events have same echo amplitude $A$. Although, this approach does not cover entire space of former searches, it makes a complementary search in overall.

A promising approach for probing quantum‐modified horizons in gravitational‐wave data uses the phenomenological “Boltzmann echoes” waveform \cite{Oshita:2019sat,Wang:2019rcf}. Its central ingredient is the Boltzmann factor $\exp[-\frac{|\omega -2 \Omega_{H}|}{2T_{H}}]$ which resembles the thermal weighting in Hawking’s radiation spectrum and governs the partial reflection of waves near the black‑hole horizon. This term is responsible to stimulated Hawking radiation making this process observable by gravitational wave detectors. Here, the Boltzmann factor originates from Hawking tunnelling rate to fuzzy states of quantum BH (please see Fig. \ref{echo_pic_3} for this waveform). Note that, repeating echoes time delay modifies this factor to $\exp[\frac{\omega-2\Omega_{H}}{2T_{H}}+i\omega\Delta t_{\rm echo}]$~\cite{Oshita:2019sat,Wang:2019rcf}.

Next section describes method and search pipeline and false/true positive estimations for GW190521. Section~\ref{Results} provides combined Bayes factor estimation and an alternative approach to combine events hierarchically assuming a target distribution. Finally, I conclude with the search results and findings.

\section{\label{Method and search pipeline} Method and search pipeline}

Adopting a novel perspective, our model and search pipeline target echo signatures in 47 LVK BBH merger events, ensuring that these events meet the criterion, guaranteeing a false alarm rate $\rm{FAR}<10^{-3} \rm{yr}^{-1}$. I perform the search for echoes on BBH signals using the GWTC-1 \cite{LIGOScientific:2018mvr}, GWTC-2 \cite{LIGOScientific:2020ibl} and GWTC-3 \cite{LIGOScientific:2021djp}. This search includes almost the bulk of all the confident observations \cite{LIGOScientific:2018mvr,LIGOScientific:2020ibl,LIGOScientific:2021djp}. The missed events are either the marginal ones or needed a high computational cost (ones with very small mass).

One such proposal to search for quantum black holes in GW data is given by phenomenological Boltzmann echoes waveform \cite{Oshita:2019sat,Wang:2019rcf}, where the general relativistic prediction for GW signal from BBH mergers $h_{\rm GR}(\omega)$ in Fourier space is modified to:
\begin{eqnarray}
    h_{\rm GR+echoes}(A,\omega) =  h_{\rm GR}(\omega) \left[ 1+ A e^{i\phi} \sum_{n=1}^\infty {\cal R}^n \right], \label{eq:Boltzmann_template} \\ {\cal R} \equiv \exp[-\frac{\hslash|\omega -m \Omega_{H}|}{2kT_{H}} + i\omega \Delta t_{\rm echo} ], \label{eq:R}
\end{eqnarray}
where $A e^{i\phi}$ quantifies their overall amplitude, while the modulus and phase of ${\cal R}$ quantify their relative damping and temporal separation, respectively.
Generally, we expect $0\lesssim A\lesssim2$ and $0< \phi\leq2\pi$ due to GR non-linearities. Here $T_{H}$ is Hawking temperature of the black hole. Furthermore, I set  the horizon mode frequency $m\times\Omega_{H}$ to m=2 for quadrupolar gravitational radiation (with the assumption that the energy in BBH ringdown and echoes are dominated by this mode) as main frequency of search pipeline. 
In this search we only keep n=2 number of echoes. Indeed, this waveform is not as perfect as GR waveform, while it helps us in future research and establishment of better waveforms. Note that the time window to look for echoes is dictated by the model in (\ref{eq:R}), where its dependence on $\Delta t_{\rm{echo}}$ and mass dependence is given by (\ref{delay}). In this equation the final mass of black hole $M_{BH}$ is dictated by GR part of the event. In this paper, we choose a conservative prior $-13\leq \log_{10}\Lambda\leq13$ for $\Delta t_{\rm{echo}}$ in (\ref{delay}). 
The most physical and accurate term in this waveform is the  Boltzmann factor ${\cal R}$ which resembles the same factor in Hawking radiation process. This term is responsible to stimulated Hawking radiation making this process observable by gravitational wave detectors.
Indeed, the remaining assumptions made in the construction of this phenomenological waveform are less accurate because achieving greater precision necessitates numerical relativity computations.

Although, there is no doubt that the $h_{\rm GR}(\omega)$ (main event GR part) exists, we want to answer whether the echoes part exists. Existence of $h_{\rm GR}(\omega)$ helps us to obtain physical prior for echo model i.e. improvements in priors for $\Delta t_{\rm echo}$ in (\ref{delay}) or $\Omega_{H}$ and $T_{H}$ variables.

I employed PyCBC inference~\cite{Biwer:2018osg} pipeline using a dynamic nested sampling algorithm, dynesty\footnote{I used 25,000 live points in each run.} \cite{Speagle_2020}. It is based on sampling the likelihood function for a hypothesis that gives a measure of existence of a signal in the data. The likelihood function that requires a power spectral density (PSD) is supposed to be compatible with the natural assumption that the background is Gaussian. For further details and the definition of the likelihood function in PyCBC Inference, please refer to Section 2.2 of~\cite{Biwer:2018osg}.
In order to combine the events, the amplitude A for all the 47 events (they are the events satisfy the criterion of a false alarm rate $\rm{FAR}<10^{-3} \rm{yr}^{-1}$) is fixed to a universal value and the individual Bayes factors of events B$_{\rm Event}$(A) are combine as follows.
\begin{equation}
    \rm Combined\ Bayes\ Factor = B(A) = \displaystyle\prod_{i=\rm Events} B_{i}(A) \label{eq4},
\end{equation}
where B$_{\rm i}$(A) is the Bayes factor of event i between the signal with echo hypothesis of amplitude A and noise.
The combined Bayes factors is shown in Fig. \ref{Fig_1}.

I have used two/three detector networks H1-L1/H1-L1-V1 (Hanford, Livingston, Virgo) depending on the event and available data for this analyse \cite{GWTC}. While we know that LIGO/Virgo detectors are not stationary and/or Gaussian, as this was shown in~\cite{Abedi:2016hgu} even for short range of data, we assume that the PSD around each event to be our safest choice. In order get a more accurate estimation of the likelihood, the PSDs of each event are obtained from data around each of the events which is (-256,256) second interval centered on merger. Note that there are some events that this range cannot be applicable due to each detector operation. So, in order to obtain the PSD around these events, this range is slightly adjusted. First I obtain the Bayes factor comparing the log likelihood to the log likelihood of the Gaussian noise. Then the combined Bayes factors of alternative models with different amplitude A are compared (here they are $h_{\rm GR+echoes}(A,\omega)$ and $h_{\rm GR+echoes}(A=0,\omega)=h_{\rm GR}(\omega)$ in (\ref{eq:Boltzmann_template})). I used class of phenomenological IMR waveform family IMRPhenomPv2~\cite{Khan:2015jqa,Hannam:2013oca} which is freely available as part of LALSuite~\cite{LALSuite}. Although the main search in Abedi et al.~\cite{Abedi:2021tti} for GW190521 has been performed with NRSur7dq4 waveform~\cite{VarmaSurrogate}, in order to make identical search with other events in this paper I employed IMRPhenomPv2 for this event. The slight change in reported Bayes factor for this event in this paper is due to change in waveform. It is worth to mention that other waveforms/changes have shown consistent result for this event~\cite{Abedi:2021tti}. Note that the details on the results and smoke tests of the analysis for single event and its posterior for other parameters can be found in~\cite{Abedi:2021tti}.

For each event I run for discrete set of amplitudes, where each run has different seed number. Since the Bayes factor estimation in PyCBC has error, it would be hard to read the result. Due to this error and in order to get smooth/stable result, for each amplitude $A$, the Bayes factor density $\bar{B}(A)$ which is the average of $B(A)$ within $(A-\Delta A /2,A+\Delta A /2)$ (see Fig.~\ref{Fig_1}) evaluated. Although the lower $\Delta A$ gives a better resolution, it leads to more fluctuations and error. In order to improve the resolution one needs to increase the number of runs (increase the amplitude bins) as well, which leads to higher computational cost. So it requires a balance between computational cost and targeted resolution. In order to get a satisfying smoothness along with optimum computational cost $\Delta A=0.2$ is chosen. In order to satisfy the approximation, three amplitude range is arranged. First range is near zero $A\sim 0$ which is the place of GR model for comparison. This range needs to have as high as possible runs to estimate the Bayes factor of GR accurately. I chose $A=(10^{-4},2\times10^{-4},3\times10^{-4},5\times10^{-4},10^{-3},2\times10^{-3},3\times10^{-3},5\times10^{-3})$ for amplitude bins to accomplish this task. Here the Bayes factor is re-normalized to $B(A\sim0)=1$. The second range is where the combined Bayes factor is $B(A)\geq 1$. This range is between $\sim(0,0.5)$ with uniform intervals of $dA=5\times10^{-3}$. In order to get satisfying result this range needs to have second priority in amplitude resolution. The last range where the Bayes factor drops significantly from 1 doesn't need to have high resolution as it already disfavoured by model when the events are combined. This range is between $(0.5,2)$ with uniform intervals of $dA=0.01$. However, for individual events one may need high resolution in all the amplitudes as well. I set 257 runs for each event and $47\times257=12079$ runs in total.

It should be noted that the combined method presented here is insensitive to weak events.
The insensitivity to weak events can be attributed to the nature of the amplitude A of the echo, which is relative to the main event signal amplitude. As events become weaker, the amplitude of the echo diminishes further into the noise. Consequently, the Bayes factor of an ideal weak event is $\rm{B(A)}_{\rm{zero\ amplitude\ event}}= 1$, making it without impact on (\ref{eq4}). This implies that the hierarchy of events is already incorporated in this combined method.

For further clarity, I summarize the data analysis steps as follows:
\begin{enumerate}
\item Perform the search for GR+echoes model (\ref{eq:Boltzmann_template}) for discrete set of fixed amplitudes $A$ within the prior range $0\leq A\leq 2$. PyCBC software was utilized, as described in the first paragraph of this section.
\item Obtain the Bayes factors $\rm B_{i}(A)$ of each run of event i and amplitude A for GR+echoes model.
\item For each event listed in Table \ref{table_1}, calculate the Bayes factor by average integration over the amplitudes: $\mathcal{B}^{\rm{GR+echo}}_{\rm{GR}}=\frac{1}{A_{max}-A_{min}}\int_{A_{min}}^{A_{max}}B(A)dA$ with $(A_{min},A_{max})=(0,2)$. The result is presented in Table \ref{table_1}.
\item To obtain the combined Bayes factor, apply equation (\ref{eq4}) using the Bayes factors obtained in step 2. The result is presented in Fig. \ref{Fig_1}.
\end{enumerate}

Please find posterior distribution of General Relativity (GR) parameters at a minimum echo amplitude of $A=10^{-4}$ in supplemental material~\cite{sup}. The retrieved parameters and their associated errors align well with the reported values for each event \cite{GWTC}.

Please note that as seen in (\ref{delay}) and (\ref{eq:R}) among the GR parameters directly impacting the redshifted final mass and final spin of the event (such as progenitor masses, redshift, and spin parameters) are the ones contributing most directly to the echo part of the waveform. On the other hand, parameters like inclination, right ascension, declination, polarization, and coalescence phase primarily assist in refining the error estimation of the aforementioned parameters, thus not directly impacting the echo part.

In addition to the primary combined Bayes factor method (\ref{eq4}), next section introduces an alternative hierarchical Bayesian approach. This method explicitly models the distribution of echo amplitudes across events using a target Gaussian distribution $\bar{\mathcal{P}}(\rm{A},\bar{A},\sigma)$. By hierarchically combining individual event Bayes factors while accounting for selection biases and parameter uncertainties, this approach infers the hyperparameters $\bar{A}$ (mean amplitude) and  $\sigma$ (standard deviation) of the underlying population. Notably, this framework generalizes the primary method: the combined Bayes factor (\ref{eq4}) emerges as the special case $\sigma\rightarrow 0$.

Data presented in this paper and configuration files to replicate the results (including priors for GR parameters) available at \cite{Search-for-echoes-on-the-edge-of-quantum-black-holes}.

\subsection{False positive and true positive estimation for GW190521}\label{PyCBCpvalue}
In this section, false/true positive probabilities for GW190521 are calculated. As Table \ref{table_1} confirms evidence in favour of echoes for this event, this section provides supplementary validation for the obtained results.

To ensure a robust outcome, I utilized two distinct approaches to obtain false and true positives. In the following subsections both of the methods are explained.

\subsubsection{Using samples from GR posterior}
\begin{figure}
\centering
\includegraphics[width=0.9\textwidth]{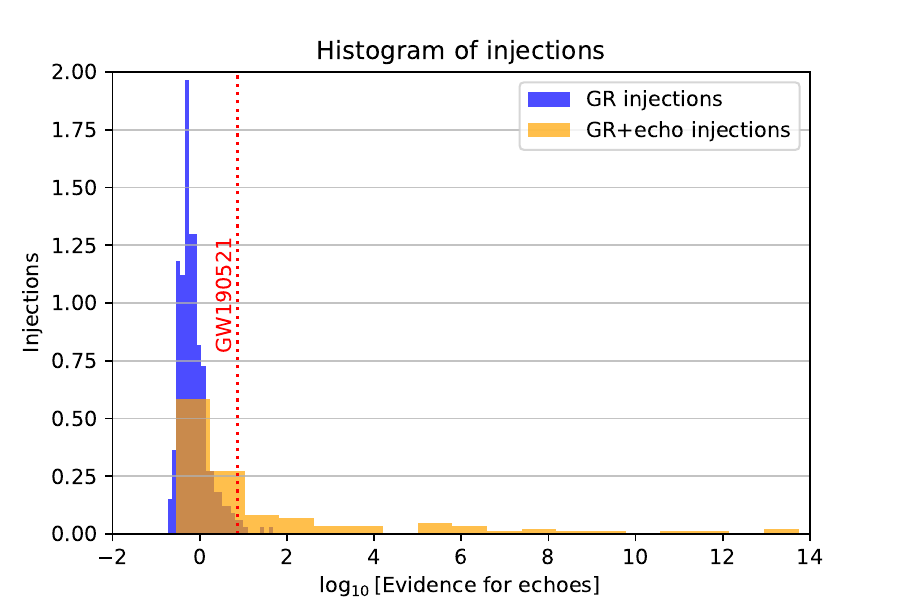}
\caption{Histograms to quantify false positive (GR injections) and true positive (GR+echo injections). Comparing to the GW190521 echo, we obtain their values as $1.46^{+1.17}_{-0.875} \%$, and $34.5\pm 7.3\%$ respectively. Note that this figure is a re‐production of the data originally presented in Fig. 4 of \cite{Abedi:2021tti}. \label{hist_pycbc}}
\end{figure}

This is the same analysis performed in \cite{Abedi:2021tti}. In this section we perform the analysis using NRSur7dq4 waveform \cite{VarmaSurrogate}. Thus, for consistency the Bayes factors of injections are compared to the one obtained with this waveform. Even though this section focuses on the outcome of the single event GW190521, it satisfies the validation test for ensuring the proper functioning of the method and model.
Since in this part we don't need to combine events and making a grid for amplitude A as the former sub-section, I made it as free parameter inside its prior range of the model $0\leq A\leq 2$. In order to evaluate the true positive we have to inject GR models containing echoes. The GR only model for injections are drawn from random samples given by GR posterior, which is obtained from 10 different seed runs. Note that in the GR run, due to the marginalization over the coalescence phase of the binary, this particular parameter was not included in the corresponding posterior sample. As a result, the random samples consist of 14 parameters, and the coalescence phase was estimated additionally using waveform fitting to the data. Once the GR part is selected out of random samples of its posterior,  the echo part is added. The echo part is within a random selection from uniform priors of $-13\leq \log_{10}\Lambda \leq13$, $0\leq A\leq 2$, and $0\leq \phi\leq 2\pi$ which is given in previous part. In order to evaluate the background and minimize non-Gaussian and/or non-stationary effects the data around the event is being used. Therefore, they are injected at random offset times of $\pm[5,32]$sec compared to GW190521 merger time. The gap $\pm5$sec around GW190521 is due to preventing any contamination of data. In order to compare the foreground (true positive) to the background (false positive), I also injected a sample with zero echo amplitude for any corresponding GR+echo injection at the same offset time.
In the next step run the same script job for injections with GR+echo and GR waveform. The number of injections performed for zero amplitude echoes (GR waveforms) is 343, while this number is 110 for non-zero amplitude echoes (GR+echo waveform). Fig. \ref{hist_pycbc} is the results of this analysis in a histogram plot (this figure is a re‐production of the data originally presented in Fig. 4 of \cite{Abedi:2021tti}). We find injection Bayes factor for GR+echo with $34.5\pm 7.3\%$ and GR with $1.46^{+1.17}_{-0.88} \%$ being larger than $\mathcal{B}= 7.5$ the echo Bayes factor for GW190521 with NRSur7dq4 waveform~\cite{Abedi:2021tti}. Then the empirical likelihood ratio $=24.3^{+42.7}_{-11.8}$ which is the ratio of the foreground probability to the background probability is obtained.

Uncertainties for these estimates (and throughout subsequent sections) are quantified via bootstrapping. To quantify sampling uncertainty in our true-positive and false-positive rates, we employ nonparametric bootstrap resampling of the injection ensembles. For each set of injections (main event and main event+echo), we generate 1,000 bootstrap replicates by resampling the original Bayes factor measurements with replacement. Within each replicate, we recompute the fraction of trials exceeding the evidence threshold $\log_{10}(B) = \log_{10}(7.5)$. The resulting bootstrap distributions provide uncertainty estimates, where we report the median along with 90\% confidence intervals from the 5th–95th percentiles. Although uncertainties are substantial (e.g. 
$34.5\% \pm 7.3\%$ vs. $1.46^{+1.17}_{-0.88}\%$), the minimum true-positive rate ($\simeq 27\%$) remains an order of magnitude above the maximum false-positive rate ($\simeq 2.6\%$), confirming robust separation.

\subsubsection{Using GW190521 signal}

In this part, we employ an alternative approach to generate foregrounds and backgrounds for injections, eliminating the need for a waveform model for the GR part to generate these injections. Specifically, we extract the data strain associated with the GW190521 main event signal and add it with or without echoes into various sections of the data surrounding the event. By employing this approach, we aim to preserve the impact of the ``main" GW190521 signal on the Bayes factor, focusing solely on injecting the echo part into nearby data, with or without echoes.

In this approach, a Gaussian window function is utilized to extract the main event signal
\begin{equation}
\exp\left(-\frac{(t-t_{\mathrm{merger}})^2}{\xi^2}\right)
\times
\left(
\begin{array}{c}
\mathsf{Hanford\ data} \\
\mathsf{Livingston\ data} \\
\mathsf{Virgo\ data}
\end{array}
\right)
\end{equation}
where $t_{\rm{merger}}=1242442967.4$ sec (GPS time).
The parameter $\xi$ should be taken in such that covers the main event signal and does't leak to $\sim1$ sec echo range. Thus $\xi = 0.5$ sec is chosen.
Then instead of GR part in former section we replace it with this extracted signal.

\begin{figure}
\centering
\includegraphics[width=0.9\textwidth]{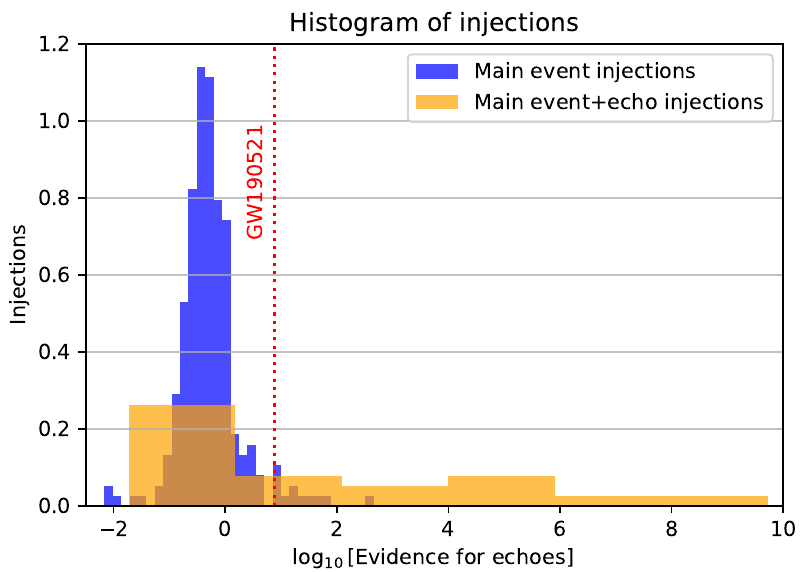}
\caption{Histograms to quantify false positive (main event injections) and true positive (main event+echo injections). Comparing to the GW190521 echo, we obtain their values as $4.4^{+2.0}_{-2.0} \%$, and $35 \pm 15 \%$ respectively.}\label{hist_pycbc2}
\end{figure}

After preparing the injections, we follow the same procedure as described earlier, running the script for injections using both the GR+echo and GR waveforms. A total of 250 injections are performed for zero-amplitude echoes (main event injections), while 20 injections are conducted for non-zero amplitude echoes (main event+echo injections). The results of this analysis are presented in Fig. \ref{hist_pycbc2} as a histogram plot. We observe that the injection Bayes factor for the main event+echo injections and for main event injections being larger than $\mathcal{B}= 7.5$ (echo Bayes factor for GW190521 with the NRSur7dq4 waveform~\cite{Abedi:2021tti}) is $35 \pm 15\%$, and $4.4^{+2.0}_{-2.0}\%$ respectively. Subsequently, we calculate the empirical likelihood ratio as $7.8^{+7.8}_{-4.0}$, representing the ratio of foreground probability to background probability.

It is important to note that here we observe a false-positive rate three times higher than that in the previous subsection. This higher rate could be due to the windowing of the main event signal, which includes unavoidable additional noise around the main event in this approach. In other words, it introduces additional noise into the analysis.

\section{\label{Results}Results}
\begin{figure}[h!]
\centering
\includegraphics[width=0.8\textwidth]{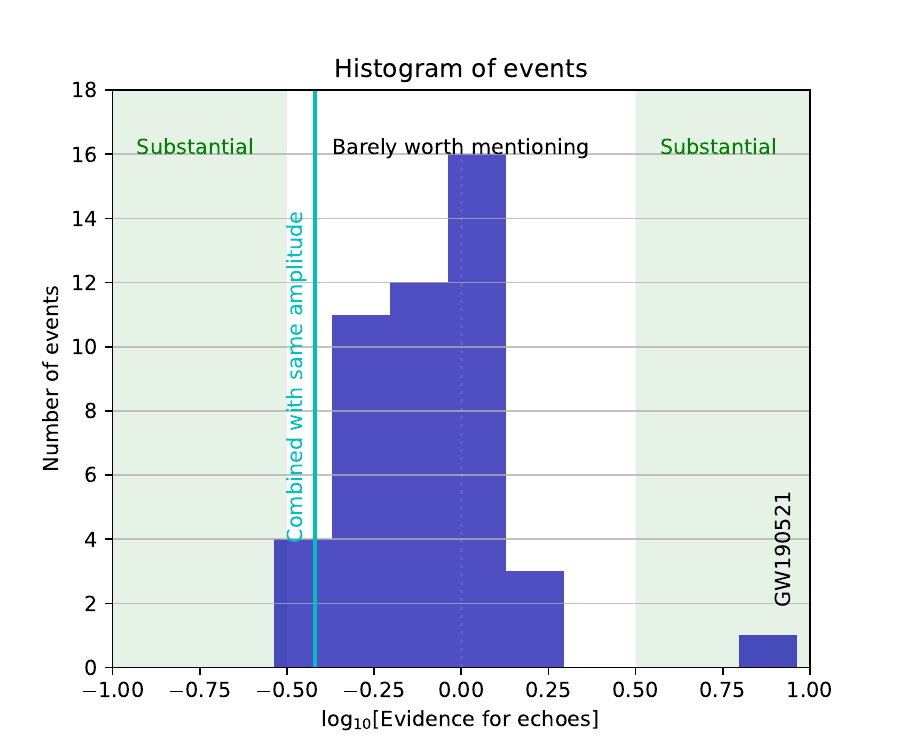}
\caption{Histogram of $\log_{10}$ Bayes factors of 47 events in Table \ref{table_1}. Vertical regions identify Jeffreys scale for interpretation of Bayes factor \cite{10.2307/2291091}.}
\label{Histogram}
\end{figure}

The result for individual events and their histogram are reported in Table~\ref{table_1} and Fig.~\ref{Histogram} respectively.
Our measured Bayes factors for 46 events (out of 47 events) lie in the range $0.3–1.6$, which falls into the “not worth more than a bare mention” regime  ($10^{-0.5} \leq \mathcal{B}^{\rm{GR+echo}}_{\rm{GR}} \leq 10^{0.5}$) on Jeffreys scale \cite{10.2307/2291091} for interpretation of Bayes factor scale. Only one event GW190521 shows a Bayes factor of 9.2 of substantial evidence $10^{0.5}<\mathcal{B}^{\rm{GR+echo}}_{\rm{GR}} \leq 10$ on Jeffreys scale.

\begin{table*}[ht]
\begin{center}
\resizebox{\textwidth}{!}{%
\begin{tabular}{|c|c||c|c||c|c|}
\hline
GWTC-1 & $\log_{10}\mathcal{B}^{\rm{GR+echo}}_{\rm{GR}}$ & GWTC-1 & $\log_{10}\mathcal{B}^{\rm{GR+echo}}_{\rm{GR}}$ & GWTC-1 & $\log_{10}\mathcal{B}^{\rm{GR+echo}}_{\rm{GR}}$ \\
\hline
GW150914 & -0.53 & GW170608 & 0.05 & GW170818 & -0.06 \\
GW151226 & -0.09 & GW170809 & 0.08 & GW170823 & -0.25 \\
GW170104 & 0.13 & GW170814 & -0.30 & & \\
\hline\hline
GWTC-2 & $\log_{10}\mathcal{B}^{\rm{GR+echo}}_{\rm{GR}}$ & GWTC-2 & $\log_{10}\mathcal{B}^{\rm{GR+echo}}_{\rm{GR}}$ & GWTC-2 & $\log_{10}\mathcal{B}^{\rm{GR+echo}}_{\rm{GR}}$ \\
\hline
GW$190408\_181802$ & -0.16 & GW190521 & 0.96 & GW$190727\_060333$ & -0.30 \\
GW190412 & -0.09 & GW$190521\_074359$ & -0.54 & GW$190728\_064510$ & -0.01 \\
GW$190421\_213856$ & 0.21 & GW$190602\_175927$ & -0.22 & GW190814 & -0.42 \\
GW$190503\_185404$ & -0.02 & GW$190630\_185205$ & -0.17 & GW$190828\_063405$ & 0.04 \\
GW$190512\_180714$ & -0.06 & GW$190706\_222641$ & -0.06 & GW$190828\_065509$ & -0.14 \\
GW$190513\_205428$ & -0.15 & GW$190707\_093326$ & -0.02 & GW$190910\_112807$ & -0.30 \\
GW$190517\_055101$ & 0.07 & GW$190708\_232457$ & -0.01 & GW$190915\_235702$ & -0.09 \\
GW$190519\_153544$ & -0.35 & GW$190720\_000836$ & -0.07 & GW$190924\_021846$ & 0.00 \\
\hline\hline
GWTC-3 & $\log_{10}\mathcal{B}^{\rm{GR+echo}}_{\rm{GR}}$ & GWTC-3 & $\log_{10}\mathcal{B}^{\rm{GR+echo}}_{\rm{GR}}$ & GWTC-3 & $\log_{10}\mathcal{B}^{\rm{GR+echo}}_{\rm{GR}}$ \\
\hline
GW$191109\_010717$ & -0.36 & GW$191222\_033537$ & -0.32 & GW$200219\_094415$ & -0.07 \\
GW$191129\_134029$ & 0.01 & GW$200112\_155838$ & -0.28 & GW$200224\_222234$ & -0.34 \\
GW$191204\_171526$ & 0.01 & GW$200129\_065458$ & -0.43 & GW$200225\_060421$ & -0.01 \\
GW$191215\_223052$ & 0.2 & GW$200202\_154313$ & 0.21 & GW$200311\_115853$ & -0.37 \\
GW$191216\_213338$ & 0.03 & GW$200208\_130117$ & 0.08 & GW$200316\_215756$ & -0.01 \\
\hline
\end{tabular}
}
\caption{Results of Bayes factor for GW echoes in GWTC-1, GWTC-2, and GWTC-3 events. Positive value of the $\log_{10}$ Bayes factor indicates a preference for the GR+echoes model over GR model, while the negative value suggests instead a preference for the GR model over the GR+echoes model. Here GW190521 shows loudest echo. Here based on \cite{10.2307/2291091} all the individual events appear as inconclusive to both GR or GR+echoes with GW190521 as exception! (see Fig.~\ref{Histogram}).}\label{table_1}
\end{center}
\begin{center}
\end{center}
\end{table*}

This search differs from previous echo analyses in two key aspects: (1) the implementation of two combined methods, discussed below, and (2) the assumption of identical echo amplitudes across all gravitational wave events.

\subsection{\label{hierarchical0}Hierarchical combination of Bayes factors}

In order to obtain the overall Bayes factor of individual events and combined events we just do $\mathcal{B}^{\rm{GR+echo}}_{\rm{GR}}=\frac{1}{A_{max}-A_{min}}\int_{A_{min}}^{A_{max}}B(A)dA$ with $(A_{min},A_{max})=(0,2)$. For combined events I got $\mathcal{B}^{\rm{GR+echo}}_{\rm{GR}}\simeq0.4$.


\subsection{\label{hierarchical}Hierarchical combination of Bayes factors assuming a target distribution}

\begin{figure*}
\begin{minipage}[l]{0.49\textwidth}
  \centering
  \includegraphics[width=1.0\linewidth]{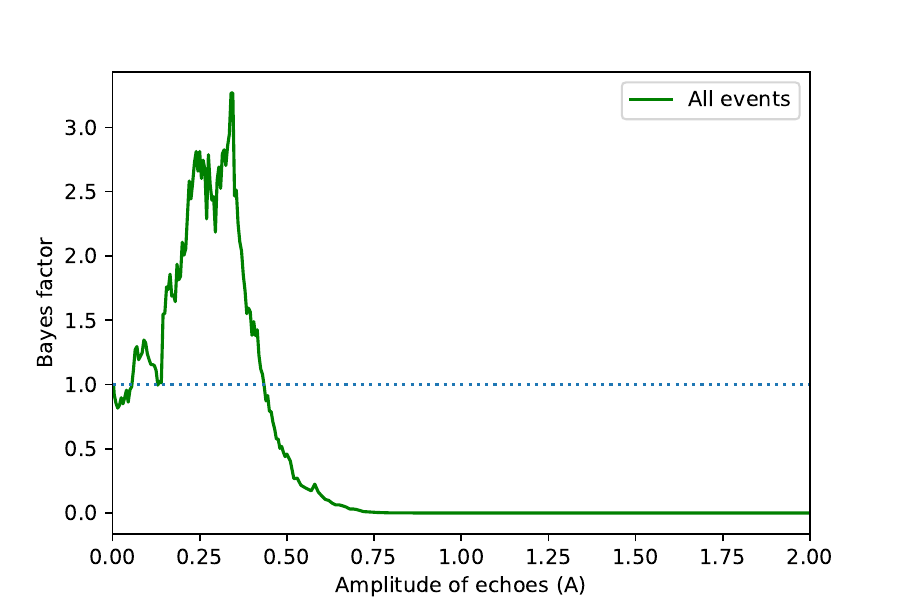}\hfill
  \label{BayesFactor}
\end{minipage}
\hfill{}
\begin{minipage}[r]{0.49\textwidth}
  \centering
  \includegraphics[width=1.0\linewidth]{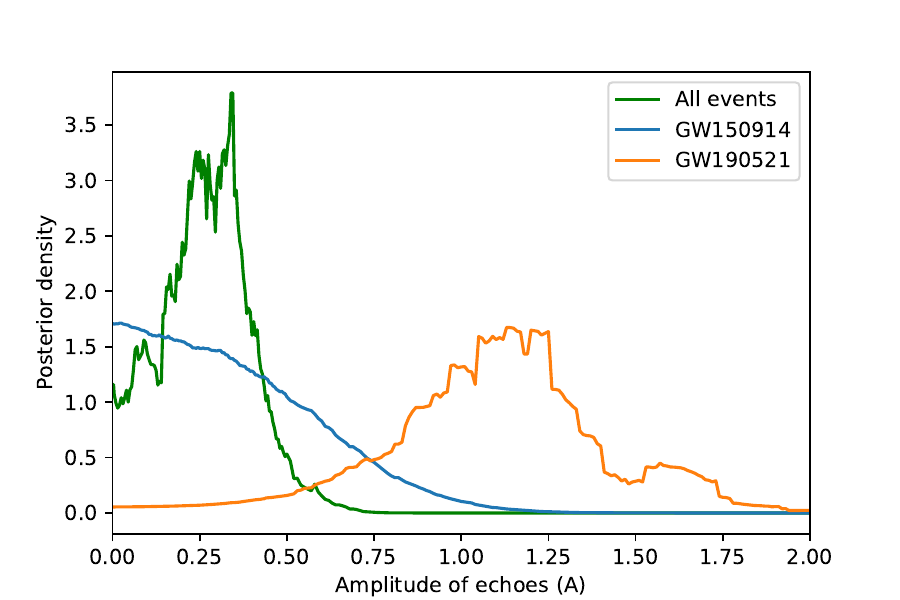}\hfill
  \label{NormalizedBayesFactor}
\end{minipage}
\caption{(a): Combined Bayes factor density in terms of amplitude for 47 events. Combined events give an overall value of $\mathcal{B}^{\rm{GR+echo}}_{\rm{GR}}\simeq0.4$ for Bayes factor. (b): Combined Bayes factor posterior density where the Bayes factor curves are normalized by $\int_{A_{min}}^{A_{max}}B(A)dA$. Here, in order to compare the individual events (GW150914 and GW190521) and combined events we plot the posterior density of the Bayes factors.}
\label{Fig_1}
\end{figure*}

\begin{figure}[h!]
\centering
\includegraphics[width=0.9\textwidth]{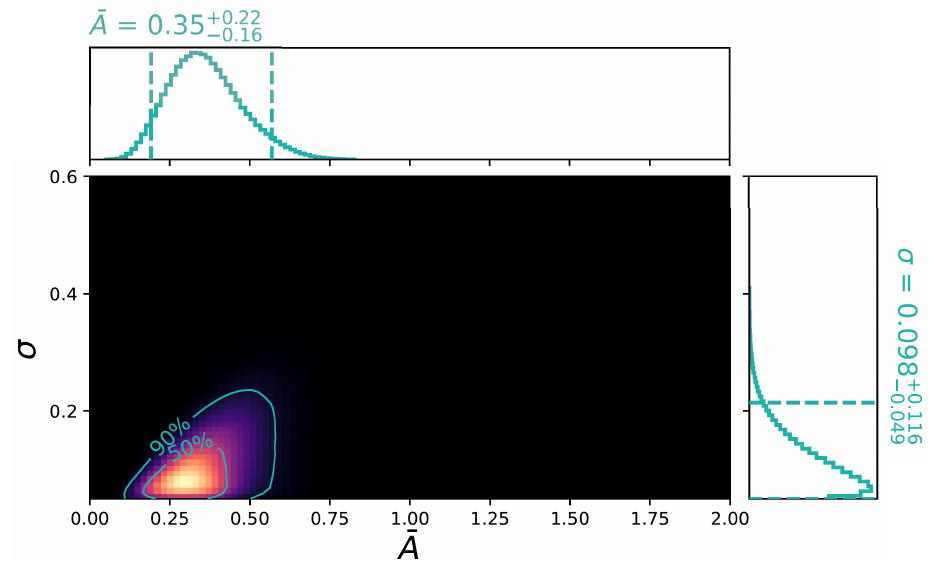}
\caption{Posterior for extracted parameters $\bar{A}$ and $\sigma$ from target distribution $\bar{\mathcal{P}}(\rm{A},\bar{A},\sigma)$ for 47 events conducted in this paper.}
\label{pycbc_posterior}
\end{figure}

This section, instead of using the hierarchical combination of Bayes factors as in the former approach, focuses on extracting distribution parameters from multiple uncertain observations that are subject to selection biases. To achieve this, I assume a target Gaussian distribution $\bar{\mathcal{P}}(\rm{A},\bar{A},\sigma) = e^{-(A-\bar{A})^{2}/2\sigma^{2}}/\sqrt{2\pi\sigma^{2}}$ with a mean of $\bar{A}$ and a standard deviation of $\sigma$ for the measured parameter in this study. Then I combine them hierarchically \cite{Isi:2019asy,Mandel:2018mve},
\begin{equation}
\mathcal{P}(\bar{\rm{A}},\sigma)=\int\prod_{i=\rm{Events}}d\rm{A}_{i} \rm{B}_{i}(A_{i}) \bar{\mathcal{P}}(A_{i},\bar{A},\sigma) \label{eq:6}
\end{equation}
Here, $\mathcal{P}(\bar{A},\sigma)$ gives a posterior function for hierarchical combination of events for parameters $\bar{A}$ and $\sigma$. With the above choice of likelihood and $\sigma\rightarrow 0$, our hierarchical approach reduces to the combined Bayes factor method described in (\ref{eq4}) for combining events. This can be seen in Fig. \ref{pycbc_posterior}, as we approach $\sigma\rightarrow 0$ limit, it tends to Fig. \ref{Fig_1} for all events.
Fig.~\ref{pycbc_posterior} presents the extracted parameters $\bar{A}$ and $\sigma$ obtained from this analysis. In this plot the lower value of 0.05 for $\sigma$ is implemented to avoid errors caused by the Bayes factor. As it can be seen in this plot, peak of $\bar{A}$ is consistent with Fig.~\ref{Fig_1} while $\sigma$ peaks near zero. This shows that there is no spread between events.

\section{\label{Conclusion}Conclusion and discussion}

I presented the outcome which gives measure of possibility for preference of $h_{\rm GR+echoes}$ over $h_{\rm GR}$  based on Bayes factors comparison of these two models. The 47 analysed events in Table \ref{table_1} and Fig. \ref{Histogram} for individual events show inconclusive result in preference for GR or GR+echo, although with slight preference for GR but not by much. The main scope and result of the paper is combination of echoes for large number of events. We see that the combined Bayes factor which is $\mathcal{B}^{\rm{GR+echo}}_{\rm{GR}}$ $\simeq 0.4$ is still inconclusive about GR+echo and GR. It is realised that this combining method gives four order of magnitude higher Bayes factor compared to when we simply combine the individual events Bayes factor via multiplication $\displaystyle\prod_{i=\rm Events} \displaystyle{\mathcal{B}_{i}}^{\rm{GR+echo}}_{\rm{GR}}=3\times10^{-5}$.  
In another words the fact that the combined Bayes factor for preference to GR has dropped from $\sim3.3\times10^4$ 
to $\sim 2.5$ indicates that there are still much to do in method improvement. Additionally, the large number of events and computational costs is a guarantee against Bayes factor hack making the result robust. 

The only event that has shown evidence for preference of GR+echo model is GW190521 with $\mathcal{B}^{\rm{GR+echo}}_{\rm{GR}}=9.2$ (see Fig.~\ref{Histogram}). This is the most massive and confidently detected BBH merger event observed to date~\cite{LIGOScientific:2020ibl}. The remarkably loud ringdown of this event  warrants special attention, making it a prime candidate for investigating gravitational wave echoes. After conducting a thorough analysis of this outlier event using two distinct methods, I found false detection probabilities of $1.46^{+1.17}_{-0.88} \%$ and $4.4^{+2.0}_{-2.0} \%$. The ratio of foreground probability to background in each case yielded an empirical likelihood ratio of $24.3^{+2.7}_{-11.8}$ and $7.8^{+7.8}_{-4.0}$. I refer the detailed interpretation and investigation about this event to Abedi et al.~\cite{Abedi:2021tti}. 

Presuming a simple speculation that we can compare all the events as same (echo model remain same for all the 47 events and their echo amplitudes compare to main event amplitude doesn't change by much despite the change in initial condition of the progenitor BBH mergers) and all the 47 BBH events should show evidence for echo signals in this model and the space of parameters considered in this search, I found an upper bound  amplitude $A< 0.4$ (at 90\% confidence level) for echoes. I remind the reader that bounds from our search only relate to the family of echo waveforms considered here.

Additionally, I employed an additional method to combine the events, wherein the parameters are extracted from multiple uncertain observations affected by selection biases. In this approach, a target distribution of Gaussian with mean $\bar{A}$ and standard deviation $\sigma$ for the search parameter A is assumed. Then I combine them hierarchically as shown in (\ref{eq:6}).  Although the extracted parameter $\bar{A}$ and its error, as depicted in Fig. \ref{pycbc_posterior}, is consistent with the peak observed for echo amplitude in Fig.~\ref{Fig_1}, the extracted parameter $\sigma$ shows that there is no significant spread between events.

It is worth to note that I didn't see any evidence for echoes in O1 in contrast to~\cite{Abedi:2016hgu,Westerweck:2017hus,Nielsen:2018lkf}, possibly because the model I used here is different and has much suppressed amplitudes in contrast to ADA model in~\cite{Abedi:2016hgu}.

It is crucial to underscore that, considering the present sensitivity levels of gravitational wave detectors \cite{LongoMicchi:2020cwm,Ma:2022xmp}, there's no concrete evidence of stimulated Hawking radiation from black hole mergers or Boltzmann echoes at $\geq5\sigma$. Taking a cautious approach, this analysis stands as the most stringent constraint on the amplitude of this signal to date.

In order to do a better search for quieter echoes, we might need to have a more physical echo waveforms. In another words, concrete models from alternatives to GR are needed to use in PyCBC pipeline.
Without better models, we might wait for O4. Observations will improve in number. LISA, Einstein Telescope, and Cosmic Explorer will make a big breakthrough in sensitivity in search for alternatives to GR.

\ack
I would like to thank Niayesh Afshordi and Alex B. Nielsen for helpful comments and discussions. I also thank Conner Dailey for suggestion about Fig.~\ref{echo_pic_3}. I thank the Max Planck Gesellschaft and the Atlas cluster computing team at AEI Hannover for support and computational help. I was supported by ROMFORSK grant Project. No. 302640. This research has made use of data, software and/or web tools obtained from the Gravitational Wave Open Science Center (https://www.gw- openscience.org), a service of LIGO Laboratory, the LIGO Scientific Collaboration and the Virgo Collaboration. LIGO is funded by the U.S. National Science Foundation. Virgo is funded by the French Centre National de Recherche Scientifique (CNRS), the Italian Instituto Nazionale della Fisica Nucleare (INFN) and the Dutch Nikhef, with contributions by Polish and Hungarian institutes.

\end{document}